\documentclass[journal,12pt,onecolumn,draftclsnofoot,]{IEEEtran}
\IEEEoverridecommandlockouts
\usepackage[T1]{fontenc}
\usepackage[pdftex]{graphicx}
\usepackage{amsmath}
\usepackage{tikz}
\usepackage{pgfplots}
\usepackage{hyperref}
\usepackage{verbatim}
\usepackage{algorithm}
\usepackage{algpseudocode}

\graphicspath{{fig}}
\pgfplotsset{width=7cm, compat=1.18}
\usetikzlibrary{positioning}
\begin{document}

\title{Fault-Tolerant Spectrum Usage Consensus for Low-Earth-Orbit Satellite Constellations} 
\author{ 
  Arman~Mollakhani and Dongning Guo\thanks{
  Department of Electrical and Computer Engineering,
  Northwestern University,   Evanston, IL.  Emails: \{arman.mollakhani, dguo\}@northwestern.edu
}
}
\date{}

\maketitle

\begin{abstract}
  Operators of low-Earth-orbit (LEO) non-geostationary satellite networks, also known as mega-constellations, are required by the current regulations to share all available satellite spectrum.  This paper proposes
  a consensus mechanism 
  to facilitate 
  spectrum sharing with accountability by 
  multiple operators,
  a 
  subset of which may even be adversarial. A distributed ledger is used to securely record and keep track of the state of
  consensus on spectrum usage, including interference incidents and the corresponding
  responsible parties.
  A key challenge is that the operators generally do not have initial agreement due to noise in their analog measurements.
  To meet this challenge, two categories of spectrum-sharing solutions are
  studied in detail. The first category employs an exact Byzantine fault tolerant (BFT) agreement model; the second category
  utilizes an approximate BFT agreement model.
  Practical considerations were taken into account regarding the BFT agreements, substantiated by numerical findings on the feasibility of the proposed solutions within the context of non-geostationary orbit satellite networks (NGSO).
\end{abstract}

\thispagestyle{empty}

\begin{IEEEkeywords}
    Non-geostationary orbit (NGSO) satellite networks, agreement protocols, automated enforcement, Byzantine fault tolerance, distributed ledger
\end{IEEEkeywords}

\section{Introduction}

Clear rules must be established for wireless operators to share spectrum and
effectively address conflicts that may arise between them. This task becomes particularly challenging when a significant number of transceivers are involved. Prominent examples of such scenarios include Citizens Broadband Radio Service (CBRS) and non-geostationary orbit (NGSO) satellite communication services. 
In the latter case, satellite operators are required by Article 9 of the International Telecommunication Union (ITU) Radio Regulations and the U.S.\ Federal Communications Commission (FCC) to coordinate in good faith to avoid harmful mutual interference in the 
satellite frequency bands
~\cite{berry2024spectrum}.

NGSO satellite constellations, commonly known as mega-constellations,
are comprised of satellites that typically orbit at several hundred kilometers (km) above ground.
The latest low Earth orbit (LEO) NGSO systems, comprising 
a constellation of satellites, are primarily designed to provide widespread broadband Internet access, ensuring comprehensive coverage 
around
the globe.
About 6,000 LEO satellites have already been deployed by 
Starlink and OneWeb, among others. 
Altogether, nearly ten operators have filed applications to launch tens of thousands of LEO satellites in total.
This paper offers techniques to
facilitate LEO satellite spectrum sharing
and contributes to a general solution to spectrum rights in outer space~\cite{berry2024spectrum}.

For sharing purposes, operators would benefit from a system that monitors and establishes the ground truth (or an approximation of it) about what frequency 
bands each operator uses in each area over time. 
For transparency, accessibility, and accountability, it is also desirable to record the operators' spectrum usage information in an 
immutable ledger.
It is often undesirable for operators to trust and rely on
a 
central monitoring and recording system.
This paper focuses on designing a decentralized system for
establishing consensus among 
multiple operators 
regarding spectrum usage. 

The challenges of building such a system encompass complexities in communication, limitations in sensor availability, and reliance on analog measurements. To establish consensus on 
spectrum usages, operators rely on their initial perception of reality obtained through analog measurements. However, analog measurements present a unique concern as the ground truth is generally unknown to any node, resulting in as many distinct perspectives or values as the number of 
sensors involved. This stands in contrast to most existing 
decentralized systems that are designed to resolve a small number of valid candidate values.
Consequently, the key lies in finding a solution that addresses the complexities of achieving consensus in the presence of an obscured ground truth while striving to approximate it as closely as possible.
 
To ensure agreement in the presence of faulty or
adversarial operators, we 
allow up to a small minority of operators to 
be Byzantine, which means that no assumption is made on which ones and how they deviate from the protocol. In this context,
agreement or consensus means all honest operators--those that abide by the protocol--reach an agreement on values that are
sufficiently close to the ground truth.
The consensus-based ledger system can be extended to enable
trading and enforcing mechanisms 
through smart contracts. 

This paper is the first to propose a decentralized consensus-based spectrum-sharing mechanism for mega-constellations.
By addressing two types of faults, namely, noise in analog measurements and adversarial operators, our proposed mechanisms enable network operators to reach consensus on spectrum usage effectively. Additionally, the designs provide a reliable method for recording consensus values in a decentralized ledger, ensuring adherence to spectrum-sharing rules and enabling further value-added applications. 
More specifically, we delve into two distinct categories of distributed consensus mechanisms designed to withstand a certain number of adversarial 
operators. The first category employs 
exact Byzantine Fault Tolerant (BFT) protocols, where operators eventually agree on a single value. The second category encompasses approximate BFT protocols, wherein honest operators output values that deviate by no more than a specified margin. 
To achieve consensus on spectrum usage across all operators and at all times, 
the spectrum and coverage area are divided into blocks, and each operator takes measurements and executes the consensus mechanism for
each time interval, 
sub-band, and area. Our research
validates the feasibility of implementing these 
consensus mechanisms on a global scale, 
determining the necessary
sensor density for operators to detect interference and 
reach consensus on resolving it.

The remainder of the paper is structured as follows: Sec.~\ref{sec:Related Work} 
surveys related research. 
We discuss interference of mega-constellations in Sec.~\ref{sec:NGSO Satellite Networks}, followed by some preliminaries in Sec.~\ref{sec:Network Model}.
In Secs.~\ref{sec:Exact Agreement} and~\ref{sec:Approximate Agreement}, we introduce schemes for reaching exact and approximate agreements, respectively.
Sec.~\ref{sec:Sequence of Tensors} explores 
methods for storing 
consensus results.
Furthermore, in Sec.~\ref{sec:Post Consensus}, we suggest some trading and enforcing mechanisms.
Sec.~\ref{sec:Practical Considerations in NGSO Networks} 
presents some practical considerations and numerical results.
Sec.~\ref{sec:Conclusion} concludes the paper.

\section{Related Work}
\label{sec:Related Work}

There have been a number of studies of consensus-based spectrum sharing in the literature.
Weiss et al.~\cite{weiss2019application} was among the first to explore the concept and different uses of blockchain technology for spectrum sharing. 
Another study~\cite{zhou2020blockchain} proposed a blockchain-based scheme for opportunistic spectrum sharing in heterogeneous 5G networks. Their approach involved human-to-human users signing contracts with a base station to share spectrum with machine-to-machine devices and receiving dedicated payments based on their contributions. 
In
~\cite{xiao2022decentralized}, a 
two-layered blockchain network was
proposed specifically for the 3.5 GHz CBRS band which addresses the functional and security requirements of a decentralized spectrum access system (SAS). 
The G-Chain layer, established at a global scale, enables regulator nodes and SAS servers to maintain a unified 
ledger on spectrum regulations and local SAS service states. The L-Chains layer, maintained by SAS servers and local stable users called witnesses, is dedicated to specific geographical zones, enabling spectrum access assignment per user request.
In another related work~\cite{li2022multi}, a consortium blockchain-based framework for operator-level dynamic spectrum sharing is introduced. Regulators oversee the process while a multi-leader multi-follower Stackelberg game is utilized to optimize spectrum pricing and resource allocation, aiming to maximize participant revenue.
In~\cite{wang2021blockchain}, an extensive survey of resource management using blockchain technologies in space-air-ground integrated networks is conducted, 
including
spectrum management and monitoring.

In contrast to the preceding and other existing works in the literature, our paper focuses on consensus protocols
and addresses the challenges of spectrum usage measurement errors and faults. 
In particular, the consensus protocols at the core of this work demonstrate a level of resilience against adversarial participants. A foundational work in this context is by Lamport et al.~\cite{lamport2019byzantine}, where they introduced {\it the Byzantine generals problem} and proposed a solution. 
The scenario involves a commanding general and lieutenant generals aiming to reach an agreement on whether to attack or retreat. The 
model
allows a certain proportion of the generals, including the commanding general, to be potentially adversarial.
As long as the fraction of honest generals exceeds a threshold, the solution guarantees that 1) they reach consensus, and 2) their consensus is the commanding general's initial proposal if the commanding general is honest.
This problem is also referred to as the Byzantine broadcast problem.
Pease et al.~\cite{pease1980reaching} extended the solution to the multi-valued agreement problem, enabling nodes to agree on values beyond simple bits.

Another relevant work is by Dolev et al.~\cite{dolev1986reaching}, who introduced the concept of approximate agreement. In this model, instead of aiming for consensus on a single value, honest nodes strive to output values that are close to each other. This approach to agreement allows for a more flexible consensus outcome, accommodating scenarios where precise consensus on a single value might be impractical or unnecessary.

Building upon the ideas from the preceding studies and protocols, 
the subsequent sections of this work  
propose mechanisms that aim to achieve robust consensus in the presence of measurement noise and adversarial participants, ensuring the reliability and integrity of the shared information.

\section{NGSO Satellite Networks}
\label{sec:NGSO Satellite Networks}
 
In 1972, the FCC adopted the Open Skies policy, which 
allowed private companies to launch and operate their own satellite systems for the first time, leading to a commercial space race.
The Iridium system in the 1990s was the first commercial NGSO satellite system to provide 
global coverage, although the success of terrestrial cellular systems rendered most satellite constellation businesses unprofitable. The tide has been changing again since
the late 2010s, when new technologies such as miniaturized satellites and mega-constellations gained significant investments and began operations worldwide.  Starlink, along with competitors like Kuiper (Amazon), ViaSat, and OneWeb, 
aimed to provide global broadband service.

The rapidly growing market has fueled the demand for clarity and more sophisticated rules governing spectrum access rights~\cite{hazlett2023open}.
In 
2023, the FCC proposed changes to rulemaking~\cite{fcc_2023} to address coordination of frequency usage, spectrum sharing, and protection of earlier-round systems through requiring NGSO fixed-satellite service (FSS) licensees and market access recipients to coordinate on their use of commonly authorized frequencies, with a sunsetting provision that would eventually transition from protection to spectrums sharing with earlier-round systems. 
According to these 
rules, if the system noise temperature of a receiver surpasses a certain threshold, each of the $n$ satellite systems involved will be limited to $1/n$ of the available spectrum~\cite{fcc_2021}.
How to detect interference, identify and verify its source, and how to negotiate and coordinate with potential interferers remain wide open questions. Meanwhile,
a satellite rights market has 
gradually emerged, where companies file claims and propose bundled agreements to settle disputes through rights assignment and auctions~\cite{hazlett2023open}. The decentralized approach proposed in this paper presents a flexible and automated infrastructure solution for such dispute resolution.

\begin{figure}
\centering
\begin{tikzpicture}
\begin{axis}[
    xlabel={Satellite density (per million km$^2$)}, 
    ylabel={Number of interference incidents},
    scaled x ticks=false,
    scaled y ticks=false,
    legend pos=north west,
    ytick={0,1000,2000,3000,4000,5000},
    legend style={font=\scriptsize},
    tick label style={font=\scriptsize},
    label style={font=\scriptsize},
    ]
    \addplot [red, dashed] table {
        2.7 110
        3.4526315789473684 182
        4.2052631578947364 266
        4.957894736842105 376
        5.710526315789473 481
        6.463157894736842 638
        7.21578947368421 794
        7.968421052631579 958
        8.721052631578947 1183
        9.473684210526315 1349
        1.0226315789473684e1 1528
        1.0978947368421052e1 1764
        1.173157894736842e1 2070
        1.248421052631579e1 2366
        1.3236842105263158e1 2674
        1.3989473684210526e1 3049
        1.4742105263157895e1 3288
        1.5494736842105263e1 3622
        1.6247368421052633e1 3990
        1.7e1 4333
    };
    \addlegendentry{4 operators, single channel}
    
    \addplot [red] table {
        2.7 10
        3.4526315789473684 21
        4.2052631578947364 28
        4.957894736842105 31
        5.710526315789473 56
        6.463157894736842 67
        7.21578947368421 90
        7.968421052631579 96
        8.721052631578947 116
        9.473684210526315 148
        1.0226315789473684e1 154
        1.0978947368421052e1 191
        1.173157894736842e1 218
        1.248421052631579e1 243
        1.3236842105263158e1 253
        1.3989473684210526e1 297
        1.4742105263157895e1 323
        1.5494736842105263e1 360
        1.6247368421052633e1 387
        1.7e1 442
    };
    \addlegendentry{4 operators, 10 sub-bands}

    \addplot [blue] table {
        2.7 41
        3.4526315789473684 65
        4.2052631578947364 93
        4.957894736842105 134
        5.710526315789473 169
        6.463157894736842 211
        7.21578947368421 286
        7.968421052631579 334
        8.721052631578947 414
        9.473684210526315 508
        1.0226315789473684e1 551
        1.0978947368421052e1 622
        1.173157894736842e1 716
        1.248421052631579e1 835
        1.3236842105263158e1 913
        1.3989473684210526e1 989
        1.4742105263157895e1 1160
        1.5494736842105263e1 1248
        1.6247368421052633e1 1415
        1.7e1 1497
        
    };
    \addlegendentry{7 operators, 10 sub-bands}

    \addplot [black] table {
        2.7 77
        3.4526315789473684 127
        4.2052631578947364 199
        4.957894736842105 259
        5.710526315789473 347
        6.463157894736842 466
        7.21578947368421 592
        7.968421052631579 695
        8.721052631578947 882
        9.473684210526315 1011
        1.0226315789473684e1 1187
        1.0978947368421052e1 1370
        1.173157894736842e1 1554
        1.248421052631579e1 1776
        1.3236842105263158e1 1942
        1.3989473684210526e1 2229
        1.4742105263157895e1 2396
        1.5494736842105263e1 2766
        1.6247368421052633e1 2963
        1.7e1 3281
    };
    \addlegendentry{10 operators, 10 sub-bands}
    
    \end{axis}
    \end{tikzpicture}
    \caption{
    The average number of interference incidents at any point in time increases with satellite density. Taking four operators with the same satellite density ranging from 3 to 17 per million km$^2$ across multiple simulations, positioned using a Poisson point process, we have each operating between 1,600 to 10,000 LEO satellites across different simulations.}
    \label{fig:satellite_interference}
\end{figure}
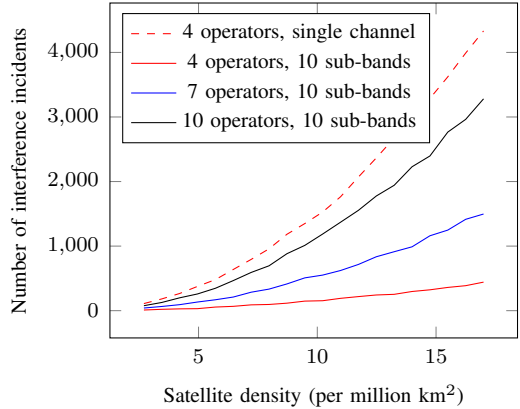

It is anticipated that,
with thousands of satellites deployed by multiple operators,
interference events become frequent unless the operators coordinate~\cite{berry2024spectrum}. We validate this concern using a stylized model where each satellite is equipped with a single direct downlink spot beam with a 3.5-degree cone angle. This implies that the spot beam is positioned directly beneath the satellite and is perpendicular to the Earth's surface. Additionally, we assume four operators with essentially the same number of satellites at the altitude of 
550 km above ground. 
An interference incident is defined as the collision of spot beams from different operators. Fig.~\ref{fig:satellite_interference} indicates hundreds to thousands of interference incidents at any point in time around the globe, where the number rises significantly with 
satellite density.
Dividing the radio band into 10 sub-bands and introducing random sub-band selection reduces incidents by approximately a factor of 10 but impacts transmission rates. These simulations exclude factors like multiple simultaneous spot beams from satellites, which would further elevate the number of interference incidents. Moreover, increasing the number of operators leads to an increase in interference events.

\section{Preliminaries}
\label{sec:Network Model}

One approach to spectrum sharing involves implementing a centralized system that monitors the activities of all operators and promptly detects interference incidents. However, this approach requires the creation of a central entity that all operators can trust, which introduces potential security and privacy risks as well as a single point of failure. To address these concerns, we propose a decentralized design, where operators collectively monitor spectrum usage, make decisions regarding interference events, and potentially participate in governance through decentralized consensus protocols. This decentralized approach empowers operators to have a more active role in decision-making and rule-making processes, fostering higher levels of trust and collaboration within the network.

\subsection{Spectrum Usage Detection}
\label{subsec:Interference Detection}

A wide range of techniques are available to individual operators for spectrum usage detection purposes.
For example, an operator may use their own transceivers (including millions of user terminals) to monitor and analyze all locations and frequency bands of interest.
In particular, one strong indicator of transmissions from other operators is sudden changes in the received signal strength or signal-to-interference-and-noise ratio (SINR). 
Various spectrum sensing techniques can be employed to detect the source of transmissions. 
In addition, participating operators may agree to incorporate uniquely identifiable pilot signals or beams in their transmissions~\cite{cabric2006spectrum}. 

In our scenario, each operator 
is capable of monitoring all operators' spectrum usage activities.  
The information an operator gathers may include physical characteristics of electromagnetic waves such as frequency, bandwidth, duration, and power or received signal strength indicator (RSSI).
Operators may employ specific equipment and techniques tailored to their own systems, and they may choose to monitor only regions where they have deployed equipment and hold a vested interest. These assumptions align with the context of mega-constellations, where satellite coverage areas are extensive and observable by most, if not all, operators~\cite{xia2019beam}.

Throughout this paper, we assume that all measurement errors are within $(-\epsilon,\epsilon)$ for the sake of simplicity, regardless of the physical characteristics of interest. This can be easily generalized to different error margins for different types of measurements made by different operators.

Dealing with transmissions from sources other than participating operators can be addressed as a separate concern. Again, 
various spectrum sensing techniques can be employed to detect the source of transmissions. 
In case no 
unauthorized transmitter
is allowed to use the spectrum,
it is beneficial for the operators to establish a pre-defined rule or agreement about intrusions.  For example, they can collectively agree that if interference from an unknown source is detected, all operators will take measures to jam the 
interferer~\cite{teng2017sharing}. This proactive approach helps mitigate the impact of unknown interference and ensures a more reliable and stable operation of the network.

\subsection{Spectrum Usage Consensus}

A major challenge lies in achieving agreement among operators based on their noisy measurements.
This challenge necessitates the development of consensus protocols and mechanisms that can reconcile differing observations and enable operators to reach consensus on spectrum usage, including the presence or absence of interference events. In general, consensus mechanisms rely on the exchange of information among operators.
While each operator gathers information from its own network of terminals and sensors, the consensus protocol can be implemented entirely within a secure computer network, where operators manage servers and exchange messages with each other.
 For the sake of simplicity, we assume each pair of operators can establish a 
 separate connection for reliable information exchange. 

We define a resource block as a sub-band within a specific region and time period. Each operator is 
uniquely identified by a number from the set $\{1,2, \dots, N\}$, 
the Earth surface is divided into $R$ regions, each 
uniquely identified by a number
from the set $\{1, 2, \dots, R\}$,
and the shared spectrum is divided into $F$ sub-bands, each uniquely identified by a number in $\{1, 2, \dots F\}$.
Additionally, time is divided into intervals of $T$ seconds.
Specifically, in each period, the goal is to reach agreement on the spectrum usage across all $R$ regions, $F$ sub-bands, and $N$ operators. Overall,
the operators need to come to an agreement on a tensor of size $R \times F \times N$ for every $T$-second interval.
Oftentimes, all operators agree to a predefined set of rules governing the default allocation of resource blocks. 
Instead of exchanging information about all resource blocks,
operators only need to reach consensus on the ones that deviate from these predefined rules, specifically interference events.
If reportable spectrum usage events are infrequent, then
the agreement protocol is only
needed for a small number of tensor elements in each time period. This significantly reduces communication costs.
This streamlined procedure simplifies the agreement process, enhances efficiency, and reduces storage and reporting requirements.

To ensure the security and reliability of the network, we consider the possibility of Byzantine operators who may deviate arbitrarily from the consensus protocol, including engaging in malicious behavior. Therefore, it is crucial to adopt agreement protocols that can tolerate Byzantine operators up to a certain threshold. The maximum tolerance is half or one-third of the total number of operators depending on the specific agreement protocol~\cite{fitzi2006optimally}.  
Agreement is reached for each consensus instance as long as a sufficient majority of participating operators follow the protocol. In fact, the design allows each consensus instance to tolerate a different set of Byzantine operators, which should include those benign operators with faulty measurements or actions within the instance.

\begin{figure}
    \centering
    \includegraphics[width=.7\columnwidth]{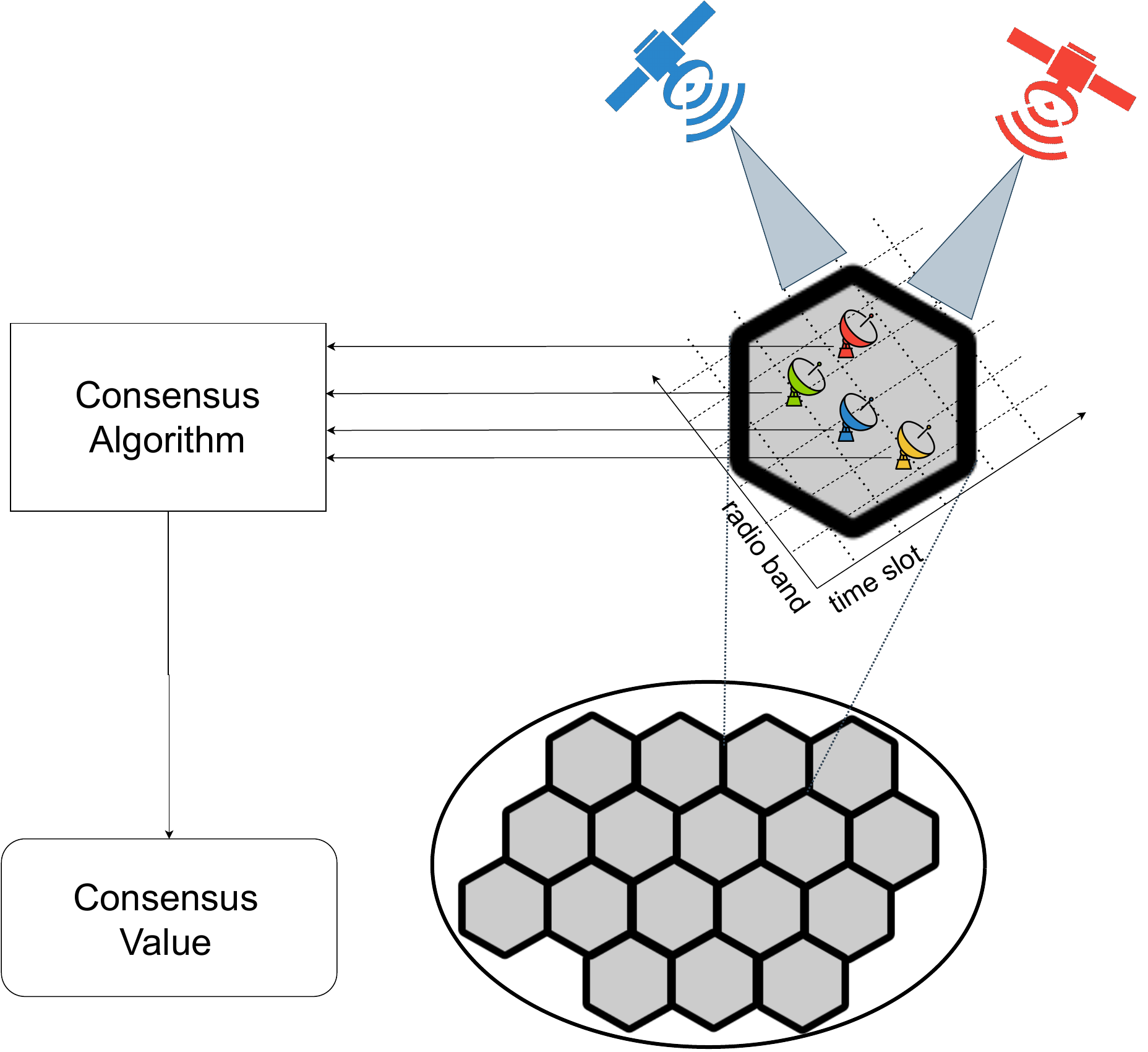}
    \caption{Consensus mechanism in an NGSO satellite network.}
    \label{fig:satellite-network}
\end{figure}

Fig.~\ref{fig:satellite-network} illustrates an application of the general proposed scheme to NGSO satellite networks. The operators measure spectrum usage within a time slot, sub-band, and region, and participate in a consensus algorithm to reach an agreement on a value (if necessary).

\section{Exact Agreement}
\label{sec:Exact Agreement}

In this section, we focus on the problem of {\em exact agreement}, namely, we consider consensus protocols for 
honest nodes to output identical values after exchanging messages about their noisy measurements. 

Each operator initializes its tensor elements based on its spectrum usage detection outcomes.
Given the requirement for agreement on all elements within a specific time period (every $T$ seconds), the agreement protocol in this context is synchronous. This means that there is a time limit 
on message delivery between operators, ensuring timely communication and coordination among them.

The objective here is for honest operators to achieve consensus on a single value based on their initial values, despite the presence of Byzantine operators.  This is known as the \textit{Byzantine agreement} (BA) problem in the BFT literature.
BA protocols fall under two main categories: deterministic and probabilistic. A deterministic BA protocol ensures agreement with certainty in a constant number of rounds. A probabilistic BA protocol is often iterative and achieves agreement with a probability that 
converges to 1~\cite{ben1983another}.

Based on the values that the tensor elements can take, we consider two cases that lead to different
agreement mechanisms.

\subsection{Binary Agreement}
\label{subsec:Binary Protocol}

We begin with a simple case where the tensor elements 
take binary values: 1 to indicate that a resource block is used by a given operator, and 0 to indicate otherwise.
The agreement problem about a tensor of
binary values is equivalent to
$R \times F \times N$ instances of
the BA problem with binary inputs.

One probabilistic approach to solving the Byzantine binary agreement problem is presented in \cite{Micali2017ByzantineA}. In this solution, it is assumed there are $N$ operators, where $N = 3f + 1$, and $f$ represents the maximum number of Byzantine operators.
As described in Algorithm~\ref{alg:binary_algorithm}, the protocol is an iterative process in which each operator takes three steps during each round until it reaches a decision on a bit and halts. 
During each step, every operator propagates its bit to all other operators. Additionally, each operator $i$ keeps track of the number of received ones and zeros, represented as $\#_i(1)$ and $\#_i(0)$, respectively.

The halting condition for operator $i$ is based on observing a supermajority of the received bits (at least $2f+1$) being equal to the bit of interest. For example, in Step 2 where the bit of interest is~1, operator $i$ halts and decides on~1 when it receives~1 from a supermajority of operators, i.e., $\#_i(1)\ge 2f+1$. (In this case, at least $f+1$ honest operators' current value is~1.) If operator $i$ does not decide to halt on Steps~1 and 2, it continues to Step 3 where it sets its bit, $b_i$, to a common 
bit produced by a random bit generator.

Several key observations can be made regarding the agreement process. Firstly, it has been demonstrated that if, at the start of Step 3, no operator has yet halted and agreement has not been reached, the probability that all honest operators share the same bit at the end of Step 3 is equal to $1/2$. Thus, on average it takes only two iterations to reach agreement. Furthermore, once agreement is established on a particular bit $b$ at any step, it will continue to hold for that same bit $b$. Lastly, if an honest operator $i$ halts at any step, agreement will be guaranteed at the end of that step.

\begin{algorithm}
  \caption{A binary agreement protocol.}
  \begin{algorithmic}
    \State \textbf{Step 1.} Each operator $i$ propagates $b_i$.
    \State \quad 1.1 If $\#_i(0) \geq 2f + 1$, then $i$ sets $b_i = 0$, outputs $out_i = 0$, and HALTS.
    \State \quad 1.2 If $\#_i (1) \geq 2f + 1$, then $i$ sets $b_i = 1$.
    \State \quad 1.3 Else, $i$ sets $b_i = 0$.
    \State \textbf{Step 2.} Each operator $i$ propagates $b_i$.
    \State \quad 2.1 If $\#_i (1) \geq 2f + 1$, then $i$ sets $b_i = 1$, outputs $out_i = 1$, and HALTS.
    \State \quad 2.2 If $\#_i (0) \geq 2f + 1$, then $i$ sets $b_i = 0$.
    \State \quad 2.3 Else, $i$ sets $b_i = 1$.
    \State \textbf{Step 3.} Each operator $i$ propagates $b_i$.
    \State \quad 3.1 If $\#_i (0) \geq 2f + 1$, then $i$ sets $b_i = 0$.
    \State \quad 3.2 Else, if $\#_i (1) \geq 2f + 1$, then $i$ sets $b_i = 1$.
    \State \quad 3.3 Else, $i$ sets $b_i$ to the common random bit generated in this round,  
    \State \quad \quad and returns to Step 1.
  \end{algorithmic}
  \label{alg:binary_algorithm}
\end{algorithm}

To mitigate predictability in common random bit generation, an effective approach, as outlined in~\cite{Micali2017ByzantineA}, utilizes cryptographic hash functions on messages signed with private keys in each round.
In this case, the probability that all honest operators generate the same bit in each round is at least 2/3, which guarantees agreement in a finite number of rounds.

There are several other ways to solve the binary agreement problem. 
In particular, running $N$ copies of a solution to the Byzantine broadcast problem described in Sec.~\ref{sec:Related Work} also solves the 
binary agreement problem
~\cite{fischer1983consensus}.
One way to solve the generals problem in a deterministic manner is proposed in~\cite{lamport2019byzantine}.

As mentioned before, each operator determines its initial value based on its own noisy measurements. This opens up the opportunity for adversaries to confuse other operators by
transmitting signals with physical features close to their decision boundary. 
In general, honest operators may initialize with different binary values. 
We call it a \textit{good consensus} when all honest operators share the same initial bit value and reach a unanimous consensus of the same bit value.

Consider a network of operators who make binary decisions by comparing the noisy RSSI measurement with a threshold $R_{th}$.
By our assumption on measurement errors, if the true RSSI is outside $(R_{th}-\epsilon,R_{th}+\epsilon)$, then all honest operators initialize with the same binary value, yielding a good consensus as long as they are a sufficient majority.
However,
if an adversary operator strategically sends signals with RSSI values within $(R_{th}-\epsilon,R_{th}+\epsilon)$, then the honest operators are likely to initialize with mixed binary decisions, preventing them from reaching a good consensus.

Evidently, if the threshold $R_{th}$ is set to be very low, then the adversary can only confuse the honest operators by transmitting at a very low power, where the benefits they can derive in terms of resource block utilization are very limited.  On the other hand, 
setting the threshold too low
would likely result in frequent false alarms, e.g., due to leakage from neighboring regions.
Therefore, it is crucial to select a threshold value that strikes a balance, being reasonably higher than the average power level of leakage waves, and at the same time lower than the power levels needed by an adversary to make a profit. 

\subsection{Arbitrary Value Agreement}
\label{subsec:Arbitrary Value Agreement}
In this subsection, we allow the tensor elements to be arbitrary numbers. 
In this context, we say a good consensus is reached when
all honest operators output the same value which falls within a given error margin of the ground truth.
One approach can be described as the following two steps: Step one is to reach agreement on the initial value of every operator. Step two is to apply a function to the vector of agreed-upon initial values so that all honest operators output the same value.

In the first step, as the network seeks consensus on an operator's initial value, it assumes the role of the commanding general in the Byzantine broadcast problem. and shares its value with others in the first round. A Byzantine broadcast protocol can be employed to achieve agreement on the initial value.
For instance, in \cite{fischer1983consensus}, an authenticated Byzantine protocol is explained in which the commanding general sends a signed message with its value to all the other operators. In the subsequent rounds, each operator records the value of valid received messages, signs them, and sends them to operators who have not signed the message yet. At the end of the protocol, if an honest operator only records a single distinct value, that value will be considered the initial value for the sending operator; otherwise, the operator concludes the commanding general is faulty and outputs a default value.
Let us denote the agreed-upon value for the initial value of operator $i$ as $v_i$.  By the end of the first step, all honest operators possess the same \textit{view} of the initial values, meaning they share a common vector consisting of the initial values of all operators, denoted as $[v_1, v_2, ..., v_N]$. 

In the second step, all operators apply an averaging function, denoted as $h$, to the view vector. This way, they collectively decide on a single value, namely $h([v_1, v_2, ..., v_N])$, which satisfies the requirements of the agreement problem and fulfills the criteria for a good consensus. One possible choice is to take the median of $v_1,\dots,v_N$.\footnote{The function proposed in \cite{dolev1986reaching} first removes the $f$ largest and $f$ smallest values and then takes the average of every $f$-th samples of the remaining values.} The use of this function ensures that the consensus value lies within the range of the initial values of the honest operators, specifically within the interval $[G-\epsilon, G+\epsilon]$ where $G$ is the ground truth value. Consequently, this exact agreement protocol guarantees a good consensus.

The preceding solution's complexity is relatively high. 
Specifically, when in need of consensus for an element of the tensor, it requires running $N$ instances of Byzantine broadcast agreement. Furthermore, each protocol run consists of $f+1$ rounds, where $f$ represents the maximum number of Byzantine nodes~\cite{pease1980reaching}. Evidently, as the number of operators increases, the complexity becomes much higher. Furthermore, the number of messages exchanged in each round is of order $O(N^2)$.
Considering at round $k$, every valid message has exactly $k$ signatures, the size of individual messages increases as the round number goes up. Therefore, as the number of operators grows, both the message complexity and size also experience a corresponding increase.

As another solution, a Byzantine protocol without authentication for arbitrary value agreement is presented in~\cite{pease1980reaching}. This protocol ensures that all honest operators compute the exact same vector of initial values, where each element corresponds to an operator. Moreover, the element corresponding to an honest operator represents its true initial value. The key concept behind this protocol is that each operator broadcasts its initial values and relays messages received from other operators regarding the messages they received from a particular operator. The protocol requires $f+1$ rounds of communication, and the number of Byzantine operators must be less than one-third of the total number of operators, i.e., $N \geq 3f+1$. The size of the messages grows in tandem with the round number in a manner akin to the authenticated protocol. As long as the number of operators 
$N$ is not large, 
the communication complexity is
low.

\section{Approximate Agreement}
\label{sec:Approximate Agreement}
In this section, we describe an alternative method that provides a trade-off between the precision of the agreement and communication complexity.
In this solution, we let operators report their noisy measurements, such as RSSI, without making a hard decision. 
We say an approximate consensus is good if it satisfies two conditions: First, the output value of each honest operator is within the range of the initial values of all honest operators.
And secondly, no two honest operators output values that differ by more than a given margin.
In particular, assuming 
the initial values of all honest operators are within $\epsilon$ of a 
ground truth,
if a good consensus is reached, every honest operator's output is also within $\epsilon$ of the ground truth.

Our objective is for operators to achieve an approximate agreement on a tensor of size $R \times F \times N$, where the corresponding elements of the tensor are within a predefined margin $\zeta$ of each other. In other words, each operator decides on a tensor that may slightly differ from the tensor of other operators. The value of $\zeta$ can vary for each element of the tensor, allowing flexibility in defining different tolerances for different tensor elements. 
One way to solve this problem is to run $R \times F \times N$ approximate agreement algorithms, each with its own margin.

Finding consensus through estimating a value within an acceptable error margin can provide notable benefits in simplicity and speed, as opposed to striving for an exact agreement. However, it is important to note that our previous consideration of Byzantine faults has been limited to the case where up to $f$ operators exhibit such faults throughout the entire protocol execution. In distributed systems, it is imperative to account for the dynamic nature of the network, where operators may encounter temporary disruptions but subsequently rejoin the network. To address this, it is advantageous to adopt a protocol that can tolerate a maximum of $f$ faulty operators at any given time, rather than exclusively considering faults throughout the entire execution. By doing so, we can enhance the protocol's resilience and effectiveness in practical scenarios.
We adopt the key techniques
proposed in~\cite{dolev1986reaching} 
to provide a protocol that reaches such approximate agreement.

In this protocol, the total number of operators denoted as $N$, must be at least three times the number of faulty operators, denoted as $f$. Therefore, the condition $N \geq 3f + 1$ ensures that a good approximate consensus is reached.
Once initialized, the algorithm consists of multiple rounds.
For simplicity, consider the value of a single element of the tensor. In
each round after collecting the responses from all other operators, each operator applies a function to multiset $V$ to get its new value. The general form of the function is given by
\begin{align}
  \label{eq:u_f,f}
  u_f(V) = \text{mean}(\text{select}_f(\text{reduce}_f(V)))
\end{align}
where $\text{reduce}_f(\cdot)$ removes the $f$ smallest and the $f$ largest values from a multiset, and $\text{select}_f(\cdot)$ chooses the smallest element and then every $f$-th element thereafter from a multiset, and mean$(\cdot)$ takes the average of a multiset.

To initialize, all operators broadcast their initial value, gather the initial values of other operators, and calculate some variables concerning the total number of rounds.
The algorithm exhibits exponential convergence. This means that in each round of execution, the range of output values for honest operators decreases by a constant factor. This shrinking factor, $c$, is given by
\begin{align}
  \label{approximate_consensus_shrinking_factor}
  c = \left\lfloor (N - 1)/f \right\rfloor - 1.
\end{align}
For a given operator, let $V$ denote the multiset of values received from all operators in a round. If an operator is halted, its halted value is used for later rounds by other operators. If no valid value is received from an operator, an arbitrary default value is picked for that operator. Furthermore, let $\delta(V)$ denote the maximum difference between all initial values known to the operator at round 1. Then the number of rounds that guarantees to reach the desired error margin $\zeta$ is
\begin{align}
  \label{approximate_consensus_round_count}
  H = \left\lceil \log_c
  ( {\delta(V)} / \zeta )
  \right\rceil.
\end{align}
Algorithm~\ref{alg:approximate_algorithm} describes the employed approximate agreement protocol in some detail.

\begin{algorithm}
  \caption{An approximate agreement protocol.}
  \begin{algorithmic}[1]
      \Procedure{{\normalfont SyncExchange}}{$v$}:
      \State Broadcast($v$)
      \State Collect $N$ responses from all operators:
          \State \hspace{\algorithmicindent} Fill in values for halted operators.
          \State \hspace{\algorithmicindent} Fill in default values, if necessary.
      \State \Return the multiset of responses
    \EndProcedure

    \Procedure{{\normalfont Main}}{$v$}:
      \State $V \gets \text{SyncExchange}(v)$
      \State $v \gets u_f(V)$
      \State $c \gets \left\lfloor (N - 1)/f \right\rfloor - 1$
      \State $H \gets \lceil \log_c\left(
      {\delta(V)}/{\zeta}\right) \rceil$

      \For{$2 \leq h \leq H$}
        \State $V \gets \text{SyncExchange}(v)$
        \State $v \gets u_f(V)$
      \EndFor

      \State Broadcast($\langle v, \text{halted} \rangle$)
      \State Output $v$
    \EndProcedure

  \end{algorithmic}
    \label{alg:approximate_algorithm}
\end{algorithm}

\section{Retrieve the Sequence of Tensors}
\label{sec:Sequence of Tensors}

Every honest operator maintains a timestamped sequence of tensor values. This sequence captures the operator's calculated tensor values along with the corresponding time period number. 
There might arise a need for a third-party client or newcomer operator to retrieve the tensor values. For instance, when a judiciary or a regulatory agency (e.g., the 
FCC) 
seeks to acquire or audit the history of consensus on spectrum usage. Additionally, the need for an accounting mechanism to enforce predefined spectrum usage rules highlights the importance of accessing the history of consensus on spectrum usage.

One apparent approach is for the third party to directly request a specific operator to share its sequence of tensors. However, this method poses a significant risk, as there is a possibility of receiving false information if the operator is Byzantine.
Two potential approaches to retrieving this history are listed below:
\begin{enumerate}
    \item
    The Aggregation Approach:
        This approach involves the third party requesting the sequence of tensors from some or all operators and then applying a predefined function to the collected set of sequences. The primary goal is to generate a consolidated sequence that closely aligns with the sequences maintained by the honest operators. By adopting this approach, the third party can effectively mitigate the influence of Byzantine operators and derive a more accurate representation of the network's historical tensor data.

    \item The Ledger Approach:
        In this method, operators collectively record a single sequence of tensors agreed upon by all honest participants. This can be accomplished through the adoption of a distributed ledger or blockchain providing a tamper-proof and transparent storage mechanism, and enhancing trust and integrity.
\end{enumerate}

The subsequent subsections delve into further detail of these two approaches, covering both exact and approximate agreement scenarios.

\subsection{Exact Agreement}
\label{subsec:exact_agreemtn}
In the scenario of exact agreement, the aggregation approach can be implemented as follows: The third party initiates the retrieval process by requesting tensor values for a specific time period $t$ from operators one by one. If the third party successfully obtains $f+1$ tensor values that are exactly identical, it conclusively determines that the retrieved tensor values represent the tensor for time period $t$.

Alternatively, to implement the ledger approach,
we let one operator acts as a proposer during each time period, suggesting a new tensor for ledger inclusion. Honest operators accept the proposed tensor only if it perfectly matches their local version for the same time period. Mismatches lead to the immediate discard of the proposed tensor. Accepted tensors are signed by operators and returned to the proposer. If the supermajority of operators (at least $2f+1$) vote in favor of the proposed tensor, it is added to the ledger. Conversely, if the supermajority rejects the proposed tensor, the proposer is identified as faulty and faces consequences. This mechanism serves as a deterrent against Byzantine operators, discouraging the submission of erroneous values.

\subsection{Approximate Agreement}
\label{subsec:approximate_agreement}
In approximate agreement, the goal is to retrieve tensor values close to those of honest operators, aiming for output values that approximate the ground truth.

The aggregation approach for retrieving tensor data can be implemented as follows: Firstly, the third party requests tensor values for a particular time period $t$ from all operators. Next, it applies the function 
defined by~\eqref{eq:u_f,f} to each element of the tensors. This ensures that the computed values fall within the output values of honest operators.  
It is possible for different clients to end up with slightly different sequences of tensors. Nonetheless, these differences are inconsequential as long as the desired function consistently produces the same outcome when applied to these tensor values.

In certain situations, there may arise a need for operators to have exact agreement on tensor values. To accomplish this, we propose storing the tensor values in a distributed ledger.
This ensures that a desired function applied to tensor values will consistently yield the same outcome for all clients.

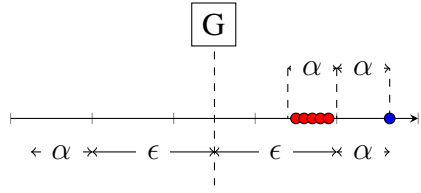
\begin{figure}
  \begin{tikzpicture}
    \begin{axis}[
        xmin=0, xmax=5,
        axis x line=bottom,
        hide y axis,
        ymin=0,ymax=5,
        xticklabels={,,},
        scatter/classes={%
            a={mark=o,draw=black}}
      ]

      \addplot[scatter,only marks,
        mark size = 2pt,
        fill = red,
        scatter src=explicit symbolic]
      table {
          3.5 0
          3.6 0
          3.7 0
          3.8 0
          3.9 0
        };

      \addplot[scatter,only marks,
        mark size = 2pt,
        fill = blue,
        scatter src=explicit symbolic]
      table {
          4.65 0
        };

      \coordinate (A) at (1,-0.5);
      \coordinate (B) at (2.5,-0.5);
      \coordinate (C) at (4,-0.5);
      \coordinate (D) at (4.65,-0.5);
      \coordinate (E) at (0.25,-0.5);
      \coordinate (F) at (2.5,1);
      \coordinate (G) at (2.5,-1);
      \coordinate (H) at (3.4,0.75);
      \coordinate (HH) at (3.4,0);
      \coordinate (I) at (4.0,0.75);
      \coordinate (II) at (4.0,0);
      \coordinate (J) at (4.65,0);
      \coordinate (JJ) at (4.65,0.75);

    \end{axis}

    \draw[<->] (A) -- (B) node[midway,fill=white] {$\epsilon$};
    \draw[<->] (B) -- (C) node[midway,fill=white] {$\epsilon$};
    \draw[<->] (E) -- (A) node[midway,fill=white] {$\alpha$};
    \draw[<->] (C) -- (D) node[midway,fill=white] {$\alpha$};
    \draw[<->] (H) -- (I) node[midway,fill=white] {$\alpha$};
    \draw[<->] (I) -- (JJ) node[midway,fill=white] {$\alpha$};
    \draw [dashed] (F) -- (G);
    \draw [dashed] (H) -- (HH);
    \draw [dashed] (I) -- (II);
    \draw [dashed] (J) -- (JJ);
    \node[draw,align=center] at (2.70,1.25) {G};
  \end{tikzpicture}
  \centering
  \caption{A Byzantine proposer.}
  \label{fig:store_approximate_values}
\end{figure}

Here we describe one possible ledger approach in some detail.
We suppose the output values of honest operators are within a range of $\alpha$. We refer to this range as the \textit{honest range}.
During each time period, our proposed approach involves one operator acting as a proposer and suggesting a new tensor for inclusion in the ledger. Honest operators 
compare the proposed tensor with their local copy. If every element of the proposed tensor falls within the specified range of $\alpha$ from their local copy, the honest operator accepts the value and casts a vote in its favor. Conversely, if the proposed tensor deviates beyond this range, it is promptly discarded.
For a proposed tensor to be added to the ledger, it must receive support from a supermajority of operators, requiring at least $2f+1$ votes in its favor. Votes here are cast by digitally signing the proposed tensor. This ensures a Byzantine proposer cannot have more than one tensor approved.
This mechanism allows Byzantine leaders to propose values that lie outside the honest range to some extent. However, there are limits to their deviation before they are held accountable for their actions. Fig.~\ref{fig:store_approximate_values} shows that the proposed value can be at most $\alpha$ away from
honest values. In this figure, the red dots are the output values of honest operators, which are
within $\alpha$ of each other. The blue dot is the value proposed by a Byzantine operator but can be at most $\alpha$ away from the honest range. Otherwise, it will be discarded by all honest operators.

Values outside the honest range may or may not be added to the ledger. Their acceptance depends on the voting outcome. Since these values are exclusively proposed by Byzantine leaders, they will be held accountable if their proposed values do not receive the supermajority vote. This accountability mechanism acts as a disincentive for Byzantine nodes to propose values that deviate from the honest range.

\section{Post Consensus}
\label{sec:Post Consensus}

The decision-making process regarding spectrum usage, as well as the associated trading and enforcement mechanisms, is executed based on the sequence of tensors stored in the ledger. To facilitate this, a smart contract may be deployed also on the ledger, responsible for validating the submitted tensors and assessing the occurrence of spectrum usage events.
In 
this section, we outline a set of ideas that can be incorporated into smart contracts for trading spectrum resources and enforcing rules, 
including payment determination and penalty imposition.
 
Through a set of smart contracts, we may let operators transfer their flexible spectrum rights to other operators, thus enabling traffic-driven dynamic and efficient resource utilization. 
We may also categorize operators into different priority levels within a region.
Once honest operators reach consensus that a lower-priority operator has violated some rules, 
the smart contract can determine appropriate punishment measures based on predefined rules in favor of higher-priority operators.

To enable seamless integration with the ledger, we may implement trading proceeds as well as rewards and penalties in the form of tokens defined on the ledger. These tokens can represent monetary units and/or resources, including radio bands or some form of usage rights. The allocation of resource blocks to operators is recorded on the ledger via  smart contracts, ensuring transparency and immutability. An operator can claim ownership of a resource block as long as the operator possesses a valid entry on the ledger. Any modifications or updates to the resource block allocation are also reflected on the ledger. 
A comprehensive exploration of the tokenization of rights to utilize assigned spectrum units can be found in~\cite{bustamante2022techno}. The authors specifically address the advanced wireless service (AWS) band (1695–1710 MHz) and present methods for temporarily transferring the spectrum rights through tokens, facilitated by the implementation of smart contracts on public or private blockchains.

We may require operators to stake a certain amount of tokens in order to share the spectrum by participating in the consensus protocol.
In the event of interference incidents, a specified amount of tokens may be deducted from the responsible operator's stake as a penalty. The severity of the penalty may increase exponentially for each interference incident within a given period. 
This exponential increment aims to discourage operators from deviating from the protocol. If an operator's stake reaches zero, its right to resource blocks may be revoked and redistributed among the remaining operators.

Prior to each consensus protocol run, operators may query the ledger to verify the validity of participants' access to the relevant resource blocks. Any operators with invalid access may be excluded from the consensus protocol. Furthermore, to incentivize operators' participation in the consensus protocol, they can be rewarded with tokens. The reward amount can be determined based on the number of valid participants in the consensus protocol. As the number of valid participants increases, the reward decreases. Such a mechanism encourages operators to deploy sensors in their respective regions of interest and actively engage in the consensus protocol. Moreover, different regions can have distinct overall rewards for each run of the agreement protocol, promoting fairness and regional participation.

By integrating these functionalities into the system, the ledger-based solution supports resource trading, ensures fair decision-making, incentivizes protocol adherence, and enables effective management of interference incidents.

\section{Practical Considerations in NGSO Networks}
\label{sec:Practical Considerations in NGSO Networks}

In this section, we 
discuss several important practical considerations
of the proposed decentralized mechanisms for
NGSO satellites.

First, effective coordination minimizes interference by Byzantine or faulty operators, leading to fewer message exchanges and reduced bandwidth usage during consensus instances. For instance, in a setting with 5 operators, 8 sub-bands, and 100,000 regions, assuming a maximum of 100 interference events in each period $T$, and each instance involves 10 rounds of 200-byte messages, an operator exchanges a maximum of $5\times100\times10\times200=1$ MB of data. This highlights efficient communication, with operators needing minimal computational power and network bandwidth. Additionally, instead of pursuing consensus on a global scale, the network can be divided into sizable regions. Within each region, a subset of operators with a shared vested interest can form an independent network to execute the consensus protocol. This targeted approach substantially reduces the number of elements in the agreement tensor, resulting in a scaled and more streamlined solution.

Detecting interference events requires the deployment of ground sensors by each provider. An interference within a cell can be detected when all honest operators have sensors in the spot beam of the interfering satellite within the cell. To achieve this, we must consider the required number of sensors for each operator to consistently detect interference events. We assume that the sensors of each provider are distributed across the Earth's surface by a Poisson point process with rate $\lambda_i$ for $i \in \{1,2,..., N\}$. Let $h$ be the constant height of all satellites across all operators and $\theta$ be half of the angle of the satellite beam cone. Assuming the projection of a satellite's spot beam onto Earth's surface to be approximately circular, the radius of this circle is approximately equal to $h \tan(\theta)$. Moreover, let $H$ denote the set of honest operators. The probability of an honest operator $i \in H$ lacking a sensor within this circular area can be determined using the corresponding Poisson distribution with rate $\lambda_i$ as
\begin{align}
    \exp  \left(-\lambda_i \pi (h \tan(\theta))^2\right) .
\end{align}
Assuming that the sensor readings of honest operators yield a good consensus, the probability of detecting an interference incident, denoted as $D$, is 
\begin{align}
    \label{eqn:detection_prob}
    P(D) &\approx \prod_{i \in H} \left[1 - \exp\left(-\lambda_i \pi (h \tan(\theta))^2\right)\right].
\end{align}

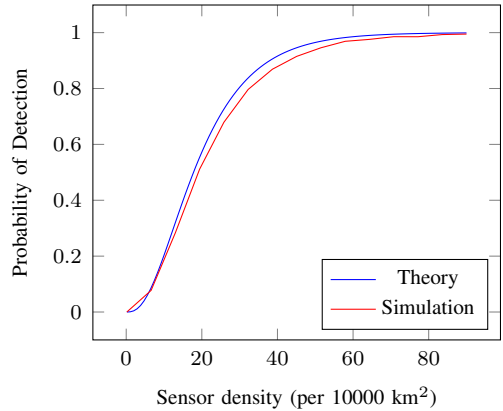
\begin{figure}
    \centering
    \begin{tikzpicture}
        \begin{axis}[
            xlabel={Sensor density (per 10000 km$^2$)},
            ylabel={Probability of Detection},
            legend pos=south east,
            ytick={0,0.2,0.4,0.6,0.8,1.0},
            legend style={font=\scriptsize},
            tick label style={font=\scriptsize},
            label style={font=\scriptsize},
        ]
        
        \addplot[blue, domain=0.17:90, samples=100] { (1 - exp(- ((x/10000) * pi * (550 * tan(deg(1.75 * pi / 180)))^2)))^3};
        
        \addplot[red] coordinates {
            (0.17, 0.0)
            (6.586428571428571, 0.0774205820810348)
            (13.002857142857144, 0.283658398030676)
            (19.419285714285714, 0.5114285714285715)
            (25.835714285714285, 0.679147465437788)
            (32.25214285714286, 0.7966069386199008)
            (38.668571428571425, 0.8690476190476191)
            (45.084999999999994, 0.9150048402710553)
            (51.50142857142857, 0.9457968809433246)
            (57.91785714285714, 0.9689629068887207)
            (64.33428571428571, 0.9758463294028148)
            (70.75071428571428, 0.9854656722126601)
            (77.16714285714285, 0.9853338479351602)
            (83.58357142857143, 0.993091537132988)
            (90.0, 0.9947996918335902)
        };
        \legend{Theory, Simulation}
        
        \end{axis}
    \end{tikzpicture}
    \caption{Probability of detection as a function of sensor density.}
    \label{fig:probability_detection}
\end{figure}

The probability described above assumes that spot beams are within the boundaries of a cell. In a series of simulations, we discretized the Earth's surface into 20,000 cells by partitioning the azimuthal coordinate into 200 bins and the polar coordinate into 100 bins. We assumed four operators, one of which is adversarial, and all of its satellites cause interference in the cell beneath them. In each simulation, operators have the same sensor density, ranging from 1 to approximately 90 sensors per 10,000 km$^2$ across simulations. Fig.~\ref{fig:probability_detection} illustrates the probability of interference detection for both the simulation and the theoretical model. As shown in the figure, with an average of around 90 sensors per 10,000 km$^2$, interference incidents are detected with a probability close to one.

NGSO satellite network operators are generally well-equipped with sensors in their coverage areas. 
Through their ground stations, they can actively monitor the spectrum usage of other operators. 
In addition, millions of user 
terminals can also contribute to the monitoring process. This approach allows operators to measure spectrum usage independently, without relying on external entities. This satisfies the requirement for the consensus protocols discussed in previous sections. 

In addition, our mechanisms offer flexibility, 
allowing operators to deploy sensors only in their regions of interest.
In situations where the proposer lacks sensors in a specific region, 
the corresponding tensor elements remain unfilled. Since all operators are aware of the sensor distribution in each region, when presented with a proposed tensor containing values for regions where the proposer lacks sensors, operators will reject the proposal. The empty elements can be populated later when 
another operator from the respective region assumes the role of the proposer.

BFT protocols perform most cost-effectively when the number of operators is not large. 
Fewer than ten operational NGSO network operators exist today~\cite{butash2021non}.
The 
number of operators 
is not anticipated to be 
large due to high barriers of entry and economies of scale.
To ensure accurate recording of spectrum usage incidents, a permissioned distributed ledger can be employed, where trusted entities participate in the creation and validation of blocks. Each operator within the network can maintain a node in the ledger infrastructure.
The solutions discussed in previous sections are well-suited to achieve consensus in spectrum usage within satellite networks.

The allowance of Byzantine operators in the network ensures robust agreements. The operators' ledgers can be used to audit spectrum usage in any regions, sub-bands, and time periods of interest.
Since agreement is needed periodically, Byzantine operators, if caught, may be excluded from participating in the network by honest operators. Thus, operators are discouraged from acting maliciously. 

Smart contracts can be utilized in the distributed ledger system to enforce incumbent mega-constellations' priority rights.
If an operator with a low priority for a spectrum resource
causes interference to 
a higher-priority operator, 
the low-priority operator's access rights can be revoked.
For instance, they may be excluded from the consensus protocol by the honest operators for a period of time. This way operators are encouraged to respect the rights of high-priority operators. In light of this, overlay networks can also be implemented on top of the operators' main network considering that the main operators have higher priorities.
In general, by leveraging smart contracts, we can establish a highly-automated, market-driven spectrum-sharing and dispute-resolution system that is not only flexible and dynamic but also upholds accountability.

It is worth noting that the integration of blockchain technologies into wireless networks has been 
implemented at scale. 
For instance, the Helium Network~\cite{helium2018}
has leveraged blockchain technology to establish a decentralized wireless communication infrastructure. 
By March 2023,  the network had amassed nearly one million onboard hotspots~\cite{helium2023}, specialized communication and mining devices operated by individuals who are incentivized with tokens for relaying users' packets to their destinations. However, it is important to consider that Helium operates on a public blockchain with a large number of hotspots~\cite{jagtap2021federated}.
In the case of satellite communications, the number of operators is much smaller,
so that consensus can be achieved with minimal computational and communication complexity, and token creation is a useful option rather than an economic necessity.

\section{Concluding Remarks}
\label{sec:Conclusion}

In this paper, we have presented decentralized consensus-based spectrum-sharing mechanisms for mega-constellations, addressing analog measurement noise and adversarial operators. BFT solutions have been used as a means to reach consensus on spectrum usage. We have utilized a distributed ledger to store consensus results, complemented by a set of smart contracts to facilitate trading spectrum resources and enforcing rules. Additionally, we have introduced methods to facilitate the retrieval of consensus values by new operators, a regulator, or clients. We have also validated the feasibility of implementing these mechanisms on a global scale and determined the necessary sensor density for operators to detect and resolve interference.

An important aspect that was not addressed in this paper pertains to the incentive compatibility of the proposed mechanisms,
i.e., whether these mechanisms align with the best interests of operators, both honest and adversarial, to determine their effectiveness. This analysis typically falls within the domain of game theory, where strategic interactions among operators are studied. 
The incentive compatibility of the proposed mechanisms can be studied from different aspects.
In the context of consensus protocols, an important consideration is whether the utilized BFT protocols are incentive compatible, meaning whether it is in the best interest of all rational nodes to follow the protocol exactly. A noteworthy contribution in this area is the introduction of an incentive-compatible Byzantine fault-tolerant model and protocol presented in~\cite{aiyer2005bar}. It introduces the Byzantine-altruistic-rational (BAR) model for constructing cooperative services and describes the implementation of BAR-B—a backup service tolerant to both Byzantine and rational users using an asynchronous replicated state machine.

Moreover, in the context of dynamic spectrum sharing,
the problem of unlicensed spectrum-sharing by strategic operators is investigated  in~\cite{teng2017sharing}. The authors propose a sharing scheme that facilitates dynamic spectrum borrowing and lending between operators. This scheme effectively addresses time-varying traffic conditions and has been proven to achieve a perfect Bayesian equilibrium, where each operator's strategy maximizes its expected utility.

It could be interesting to incorporate
the preceding incentive-compatible schemes 
into smart contracts to facilitate dynamic spectrum sharing among operators.

\section*{Acknowledgement}
This work was supported in part by the National Science Foundation (NSF) (grant No.~1910168) and the SpectrumX Center (NSF grant No.~2132700).

\bibliographystyle{IEEEtran}
\bibliography{refs}

\begin{thebibliography}{10}
\providecommand{\url}[1]{#1}
\csname url@samestyle\endcsname
\providecommand{\newblock}{\relax}
\providecommand{\bibinfo}[2]{#2}
\providecommand{\BIBentrySTDinterwordspacing}{\spaceskip=0pt\relax}
\providecommand{\BIBentryALTinterwordstretchfactor}{4}
\providecommand{\BIBentryALTinterwordspacing}{\spaceskip=\fontdimen2\font plus
\BIBentryALTinterwordstretchfactor\fontdimen3\font minus \fontdimen4\font\relax}
\providecommand{\BIBforeignlanguage}[2]{{%
\expandafter\ifx\csname l@#1\endcsname\relax
\typeout{** WARNING: IEEEtran.bst: No hyphenation pattern has been}%
\typeout{** loaded for the language `#1'. Using the pattern for}%
\typeout{** the default language instead.}%
\else
\language=\csname l@#1\endcsname
\fi
#2}}
\providecommand{\BIBdecl}{\relax}
\BIBdecl

\bibitem{berry2024spectrum}
R.~A. Berry, P.~Bustamante, D.~Guo, T.~W. Hazlett, M.~L. Honig, W.~Lohmeyer, I.~Murtazashvili, S.~Palo, and M.~B.~H. Weiss, ``Spectrum rights in outer space: Interference management for low {Earth} orbit {(LEO)} broadband constellations,'' \emph{Journal of Information Policy, under revision. Available at SSRN 4178793}, 2024.

\bibitem{weiss2019application}
M.~B. Weiss, K.~Werbach, D.~C. Sicker, and C.~E.~C. Bastidas, ``On the application of blockchains to spectrum management,'' \emph{IEEE Transactions on Cognitive Communications and Networking}, vol.~5, no.~2, pp. 193--205, 2019.

\bibitem{zhou2020blockchain}
Z.~Zhou, X.~Chen, Y.~Zhang, and S.~Mumtaz, ``Blockchain-empowered secure spectrum sharing for {5G} heterogeneous networks,'' \emph{IEEE Network}, vol.~34, no.~1, pp. 24--31, 2020.

\bibitem{xiao2022decentralized}
Y.~Xiao, S.~Shi, W.~Lou, C.~Wang, X.~Li, N.~Zhang, Y.~T. Hou, and J.~H. Reed, ``Decentralized spectrum access system: Vision, challenges, and a blockchain solution,'' \emph{IEEE Wireless Communications}, vol.~29, no.~1, pp. 220--228, 2022.

\bibitem{li2022multi}
Z.~Li, W.~Wang, Q.~Wu, and X.~Wang, ``Multi-operator dynamic spectrum sharing for wireless communications: a consortium blockchain enabled framework,'' \emph{IEEE Transactions on Cognitive Communications and Networking}, vol.~9, no.~1, pp. 3--15, 2022.

\bibitem{wang2021blockchain}
Y.~Wang, Z.~Su, J.~Ni, N.~Zhang, and X.~Shen, ``Blockchain-empowered space-air-ground integrated networks: Opportunities, challenges, and solutions,'' \emph{IEEE Communications Surveys \& Tutorials}, vol.~24, no.~1, pp. 160--209, 2021.

\bibitem{lamport2019byzantine}
L.~Lamport, R.~Shostak, and M.~Pease, ``The {Byzantine} generals problem,'' in \emph{Concurrency: the works of Leslie Lamport}, 2019, pp. 203--226.

\bibitem{pease1980reaching}
M.~Pease, R.~Shostak, and L.~Lamport, ``Reaching agreement in the presence of faults,'' \emph{Journal of the ACM}, vol.~27, no.~2, pp. 228--234, 1980.

\bibitem{dolev1986reaching}
D.~Dolev, N.~A. Lynch, S.~S. Pinter, E.~W. Stark, and W.~E. Weihl, ``Reaching approximate agreement in the presence of faults,'' \emph{Journal of the ACM}, vol.~33, no.~3, pp. 499--516, 1986.

\bibitem{hazlett2023open}
T.~Hazlett, D.~Guo, and M.~Honig, ``From “open skies” to traffic jams in 12 ghz: A short history of satellite radio spectrum,'' \emph{Journal of Law \& Innovation}, vol.~6, no.~1, pp. 66--94, 2023.

\bibitem{fcc_2023}
{Federal Communications Commission}, ``{Order and Notice of Proposed Rulemaking, FCC 23-29},'' Apr. 2023.

\bibitem{fcc_2021}
------, ``{Order and Notice of Proposed Rulemaking, FCC-21-123},'' Dec. 2021.

\bibitem{cabric2006spectrum}
D.~Cabric, A.~Tkachenko, and R.~W. Brodersen, ``Spectrum sensing measurements of pilot, energy, and collaborative detection,'' in \emph{IEEE Military Communications Conference}, 2006, pp. 1--7.

\bibitem{xia2019beam}
S.~Xia, Q.~Jiang, C.~Zou, and G.~Li, ``Beam coverage comparison of {LEO} satellite systems based on user diversification,'' \emph{IEEE Access}, vol.~7, pp. 181\,656--181\,667, 2019.

\bibitem{teng2017sharing}
F.~Teng, D.~Guo, and M.~L. Honig, ``Sharing of unlicensed spectrum by strategic operators,'' \emph{IEEE Journal on Selected Areas in Communications}, vol.~35, no.~3, pp. 668--679, 2017.

\bibitem{fitzi2006optimally}
M.~Fitzi and M.~Hirt, ``Optimally efficient multi-valued {Byzantine} agreement,'' in \emph{Proceedings of the twenty-fifth annual ACM symposium on Principles of distributed computing}, 2006, pp. 163--168.

\bibitem{ben1983another}
M.~Ben-Or, ``Another advantage of free choice (extended abstract) completely asynchronous agreement protocols,'' in \emph{Proceedings of the second annual ACM symposium on Principles of distributed computing}, 1983, pp. 27--30.

\bibitem{Micali2017ByzantineA}
S.~Micali, ``Fast and furious {Byzantine} agreement,'' in \emph{8th Innovation in Theoretical Computer Science, ITCS}, January 2017, single-page abstract. Full version available at \url{https://people.csail.mit.edu/silvio/Selected%20Scientific%20Papers/Distributed%20Computation/}, with title “Byzantine Agreement, Made Trivial”.

\bibitem{fischer1983consensus}
M.~J. Fischer, ``The consensus problem in unreliable distributed systems (a brief survey),'' in \emph{Foundations of Computation Theory: Proceedings of the 1983 International FCT-Conference Borgholm, Sweden, August 21--27, 1983 4}.\hskip 1em plus 0.5em minus 0.4em\relax Springer, 1983, pp. 127--140.

\bibitem{bustamante2022techno}
P.~J. Bustamante, M.~M. Gomez, M.~B.~H. Weiss, I.~Murtazashvili, and A.~Palida, ``A techno-economic study of spectrum sharing with blockchain and smart contracts,'' \emph{IEEE Communications Magazine}, vol.~61, no.~2, pp. 58--63, 2023.

\bibitem{butash2021non}
T.~Butash, P.~Garland, and B.~Evans, ``Non-geostationary satellite orbit communications satellite constellations history,'' \emph{International Journal of Satellite Communications and Networking}, vol.~39, no.~1, pp. 1--5, 2021.

\bibitem{helium2018}
A.~Haleem, A.~Allen, A.~Thompson, M.~Nijdam, and R.~Garg, ``Helium: A decentralized wireless network,'' White paper, Helium Systems, Inc., 2018, release 0.4.2 (2018-11-14).

\bibitem{helium2023}
\BIBentryALTinterwordspacing
``\BIBforeignlanguage{en}{Helium foundation protocol report 2023},'' 2023, accessed on: July 10, 2023. [Online]. Available: \url{https://web.archive.org/web/20230627083359/https://www.helium.foundation/protocol-report}
\BIBentrySTDinterwordspacing

\bibitem{jagtap2021federated}
D.~Jagtap, A.~Yen, H.~Wu, A.~Schulman, and P.~Pannuto, ``Federated infrastructure: usage, patterns, and insights from" the people's network",'' in \emph{Proceedings of the 21st ACM Internet Measurement Conference}, 2021, pp. 22--36.

\bibitem{aiyer2005bar}
A.~S. Aiyer, L.~Alvisi, A.~Clement, M.~Dahlin, J.-P. Martin, and C.~Porth, ``{BAR} fault tolerance for cooperative services,'' in \emph{Proceedings of the twentieth ACM symposium on Operating systems principles}, 2005, pp. 45--58.

\end{thebibliography}

\end{document}